\documentclass[useAMS,usenatbib]{mnras}
\usepackage{graphicx}
\usepackage[section]{placeins}

\usepackage{yfonts}
\usepackage{times} 
\usepackage{amsmath}
\usepackage{amssymb}
\usepackage{tikz}
\usepackage{xcolor}

\def\icarus{Icarus}

\def\aj{Astron.\ J. }
\def\apj{Astrophys.\ J. }
\def\apjl{Astrophys.\ J. }
\def\apjs{Astrophys.\ J.\ (Supp.) }
\def\aap{Astron.\ Astrophys. }

\def\bssA_{Bull.\ Seismol.\ Soc.\ Am. }

\def\gcA_{Geochim.\ Cosmochim.\ Acta }

\def\mnras{Mon.\ Not.\ R.\ Astron.\ Soc. }

\def\planss{Plan.\ Space Sci. }
\def\prA_{Phys.\ Rev.\ A }

\begin{document}

\title[Origin of high-inclination Centaurs]{An interstellar origin for high-inclination Centaurs}
\author[F. Namouni and M. H. M. Morais]{F. Namouni$^{1}$\thanks{E-mail:namouni@obs-nice.fr (FN) ; helena.morais@rc.unesp.br (MHMM)} 
and  M. H. M. Morais$^{2}$\footnotemark[1]\\
$^{1}$Universit\'e C\^ote d'Azur, CNRS, Observatoire de la C\^ote d'Azur, CS 34229, 06304 Nice, France\\
$^{2}$Instituto de Geoci\^encias e Ci\^encias Exatas, Universidade Estadual Paulista (UNESP), Av. 24-A, 1515 13506-900 Rio Claro, SP, Brazil}

\date{Accepted 2020 March 7. Received 2020 March 6; in original form 2019 September 30.}

\maketitle

\begin{abstract}
{We investigate the {possible}  origins of real high-inclination Centaurs and trans-neptunian objects using a high-resolution statistical search for stable orbits that simulates their evolution back in time to the epoch when planet formation ended 4.5 billion years in the past. The simulation is a precise orbit determination method that does not involve ad hoc initial conditions or assumptions such as those found in planetesimal disk relaxation models upon which their conclusions depend. It can therefore be used to independently test origin theories based on relaxation models  by examining the past orbits of  specific real objects. {Here, we examined 17 multiple-opposition high-inclination Centaurs and the two polar trans-neptunian objects 2008 KV42 and (471325) 2011 KT19. The statistical distributions show that their orbits were nearly polar 4.5\,Gyr in the past, and were located in the scattered disk and inner Oort cloud regions. Early polar inclinations cannot be accounted for by current Solar system formation theory as the early planetesimal system must have been nearly flat in order to explain the low-inclination asteroid and Kuiper belts. Furthermore, the early scattered disk and inner Oort cloud regions are believed to have been devoid of Solar system material as the planetesimal disk could not have extended far beyond  Neptune's current orbit in order to halt the planet's outward migration. The nearly polar orbits of high-inclination Centaurs 4.5\,Gyr in the past therefore indicate their probable early capture from the interstellar medium.}}
\end{abstract}


\section{Introduction}

Centaurs are some of the most intriguing objects in the Solar system. With perihelia and mean orbits inside the giant planets' domain, they are subjected to some of the strongest interactions in the Solar system.  With moderate to high eccentricities, Centaurs' orbits may be inclined by a few degrees with respect to the Solar system's invariable plane to almost 180$^\circ$ resulting in retrograde motion. Their orbital features are often taken as a sign of their violent past in the Solar system, a notion  reinforced by their so-called instability. If a Centaur orbit is integrated forward or backward in time, it will invariably either hit the Sun, the planets or be ejected from the Solar system.  As the gravitational dynamics of $N$-body systems is time-reversible, the values of the future and past lifetimes are statistically similar  \citep{Horner04,Port18} and range from 1 to 100\,Myr \citep{TiscarenoMalhotra03,Disisto07,BaileyMalhotra09,VolkMalhotra13}. 

{The exact meaning of this `instability' is, however, unclear. The future lifetime is simple to interpret literally as the Centaurs are simply ejected from the Solar system  by their encounters with the planets or experience a planetary or solar collision. This should occur in the near future in view of the 1 to 100 Myr  estimates of the future lifetime \citep{Peixinho19}. The past lifetime is trickier to interpret literally because it would mean the Centaurs cannot have lived in the Solar system more than 1 to 100\,Myr in the past. This indicates that they must all have been captured from the interstellar medium in the recent past.\footnote{In time-backward integrations, collisions are unphysical. Time-backward ejection means time-forward capture.} Conventional models that study the origin of outer Solar system objects do not advocate this idea as they rely on the relaxation of the flat planetesimal disk present 4.6\,Gyr in the past to explain not only Centaurs but all known outer Solar system bodies  \citep{Levison97,TiscarenoMalhotra03,Emelyanenko05,Disisto07,Brasser12,Nesvorny17,Fernandez18,Kaib19}. By propagating the evolution of the early flat planetesimal disk 4.6\,Gyr in the future, all surviving objects, including Centaurs,  share the same age with the Solar system. 

So are the Centaurs' real past lifetimes short or can they be as long as the Solar system's age? This question may be answered if the literal interpretation of the dynamical lifetime is applied consistently to both past and  future. This interpretation would therefore require the Centaur to reach the Solar system from the interstellar medium around say 10\,Myr in the recent past only to leave the Solar system 10\,Myr in the near future as the past and future lifetimes are similar. This would imply that the Solar system happens to be crossing the Centaur's path in the Galaxy  around the current epoch and that its capture process from the interstellar medium should be fine-tuned to yield a stay of 20\,Myr  centered on our current epoch.  Making our current epoch a privileged time in the Solar system's history to observe each known Centaur in the middle of its short Solar system stay violates the Copernican principle. 
Therefore, the past lifetime cannot be an indication of the actual age of a real Centaur and the physical instability of its orbit.  It is probably  a measure of the chaotic nature of motion at a Centaur's location. Consequently, the Centaur's age can be arbitrarily large.  Planetesimal disk relaxation models are therefore justified is assuming that Centaurs may have the age of the Solar system regardless of their short past dynamical lifetimes.

To understand why no real\footnote{As opposed to the theoretical objects that represent Centaurs in planetesimal disk relaxation models whose orbits are stable over 4.6Gyr.} Centaur orbit was known to be stable over the age of the Solar system,} it is necessary to know precisely how the dynamical evolution of real Centaurs is studied.   We will only speak of Centaurs that were observed at multiple oppositions as their orbits are known with a reasonable degree of certainty. In order to assess the stability of a real Centaur's orbit, a clone swarm is replicated from its nominal orbit consistent with the observational error bars, and their equations of motion are integrated numerically. The future and past dynamical lifetimes correspond to  the median clone survival time  in the future and in the past respectively. Two parameters enter the evolution simulation:  the integration timespan and the clone number. The first is usually shorter than the age of the Solar system because the median lifetime inferred from the first trials itself is short. The clone number usually ranges  from $10^2$ to $10^3$ owing to the computational cost of integrating larger sets but also to the acquired idea that real Centaur orbits are unstable.  However as the orbital element space of a Centaur's orbit is six-dimensional (3 positions and 3 velocities), clone sets of $10^2$ and $10^3$ correspond respectively to grids of approximately 2 points and 3 points within the error bar of each orbital element. These resolutions are insufficient to probe  the Centaurs' stable orbits as they reside in one of the most chaotic regions of the Solar system.\footnote{In mathematical terms, low resolution searches cannot identify the sticky tori of the dynamical system known to persist in the chaotic sea even in the case of high non-linearity \citep{zaslavsky07}.}  As a result, the dynamical integration of real Centaurs' motion in past studies is systematically unstable and remains in contradiction with the theoretical stable Centaur orbits found in disk relaxation models. 

Progress was recently achieved in the study of Centaur origin in the outer Solar system with the development a high-resolution statistical search for stable orbits that led to the identification of the interstellar origin of Jupiter's retrograde co-orbital asteroid (514107) Ka`epaoka`awela (\cite{NamouniMorais18b} hereafter Paper I). A swarm of a million clones was integrated backward in time for 4.5\,Gyr to the epoch when planet formation ended. About half the number of clones collided with the Sun whereas the other half were ejected from the Solar system. {The median clone lifetime was 6\,Myr.  Of the million clone swarm, 46  members survived the age of the Solar system  demonstrating that a real Centaur can be as old as the Solar system thus satisfying the Copernican principle and proving for the first time that  the past dynamical lifetime is an indication of the strength of non-linearity in the neighborhood of a Centaur's orbit. The 46 stable clones may constitute a small set with respect to the original million. However, the million sample is equivalent to a grid of just 10 points within the error bar of each orbital parameter. The low stable orbit rate therefore reflects the intrinsic limited resolution in each orbital parameter. It should also be noted that the existence of Centaur stable orbits over the age of the Solar system does not preclude a possible interstellar origin as is shown by asteroid Ka`epaoka`awela's example.  }

The high-resolution statistical orbit search applied to specific real objects is new and was used for the first time in the study of asteroid Ka`epaoka`awela. It is not a model of Centaur origin but a precise orbit determination method that it is not based on ad hoc initial conditions such as the primordial planetesimal disk's attributes that are used in typical relaxation models in the literature {\citep{Levison97,TiscarenoMalhotra03,Emelyanenko05,Disisto07,Brasser12,Nesvorny17,Fernandez18,Kaib19}. Provided the number of clones is sufficiently large, the method has, in effect, no parameter that can be modified to fine-tune its outcome. The method inputs the real asteroid's nominal orbit and the covariance matrix that represents the  observational uncertainty and yields a probability distribution of the possible orbits that the asteroid had at the end of planet formation.  The method may therefore be used on real objects to   test planetesimal disk relaxation models and the ad hoc assumptions they make about the initial state of the Solar system. }

{In this work, we study the possible origins of high-inclination Centaurs and independently test whether they originate from the flat planetesimal disk of the conventional theory of Solar system formation  \citep{Pfalzner15}. We apply the high-resolution statistical orbit search to  multiple-opposition high-inclination Centaurs and the two trans-neptunian objects (TNOs) with polar orbits.}  We call high-inclination Centaurs those with orbital inclinations from 60$^\circ$ to $180^\circ$ encompassing prograde,  polar and retrograde motion.  We limit our study to high inclinations for three reasons: first, previous work has shown that resonance capture, an essential dynamical process for all Centaurs, is more efficient for  nearly-polar and retrograde motion than for prograde motion thereby extending high-inclination Centaur lifetimes in the Solar system \citep{NamouniMorais15,NamouniMorais17,MoraisNamouni17b}. 

{Secondly, if high-inclination Centaurs originate in the flat planetesimal disk,  most of their clones will have low inclinations allowing us  to confirm their evolution within the conventional view of Solar system formation. If however such objects retain  high inclinations over the age of the Solar system, then the interstellar medium is a more probable origin}. 
It is generally believed that when planet formation ended 4.5\,Gyr in the past, the planets and the remaining compact and thin planetesimal disk that extended no further than Neptune's current orbit \citep{Gomes04}, were nearly co-planar in order to explain the low-inclination reservoirs such as the asteroid and Kuiper belts through the gravitational relaxation of the system formed by the disk and the planets.  \citep{Pfalzner15} 

Thirdly, the high-resolution statistical search applied to 20 asteroids requires about $2\times 10^7$ clones implying a significant computational volume. For comparison, however, we also simulated the evolution of (2060) Chiron, a low inclination Centaur, that helps us illustrate the dependence of the stable orbit rate on the orbit's uncertainty level.

The paper is organized as follows. In Section 2, we present the Centaur and TNO sample and describe the simulation's setting. In Section 3, we discuss the statistics of stable and unstable motion and show that 4.5\,Gyr-stable orbits exist for all objects in the study and originate in the scattered disk and inner Oort cloud regions. In Section 4, we discuss the Centaurs' interstellar origin and the possible dynamical identification of common capture events from the interstellar medium in the early Solar system.

\begin{figure}
\begin{center}
{ 
\includegraphics[width=80mm]{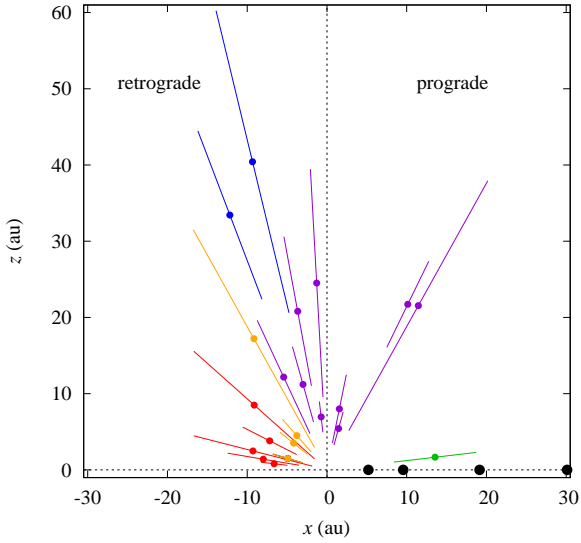}
}
\caption{Distribution of the Centaur and TNO sample. Each segment represents the aphelion and perihelion excursions centered around the object's semi-major axis (filled circle) and is inclined with respect to the planets' reference plane. The polar Centaurs are purple, TNOs  are blue, coorbital objects are orange, retrograde Centaurs are red, (2060) Chiron is green and the four giant planets are black. }\label{f1}
\end{center}
\end{figure}

\section{Object sample and simulation method}
The IAU Minor Planet Center\footnote{http://www.minorplanetcenter.net} lists 17 multiple-opposition Centaurs with inclinations larger than $60^\circ$ and perihelia larger than  3 au as of 1 December 2018. The perihelion criterion is to limit the effect of the inner planets on the Centaurs' evolution. For convenience, Centaurs are divided into 3 groups. The first is the polar Centaurs group whose orbital planes are located within $30^\circ$ from the direction of the Solar system's total angular momentum. 
In order of increasing inclination, this group includes 2010 FH92, 2007 BP102, 2010 CR140, (144908) 2004 YH32, (518151) 2016 FH13, 2014 JJ57, 2011 MM4, (342842) 2008 YB3 and 2016 LS. The second group is that of retrograde Centaurs that were studied in \cite{MoraisNamouni13b} to show that Jupiter and Saturn temporarily capture Centaurs in retrograde mean motion resonance. The objects are  2009 QY6, 1999 LE31, 2006 BZ8, (330759) 2008 SO218 and (434620) 2005 VD\footnote{Although 2009 QY6 and 2006 BZ8 have perihelia near 2 au, they were included in the study as they had demonstrated retrograde resonance capture with Jupiter and Saturn in our earlier work.}. The third group is that of Centaurs located in or near a planet's co-orbital region; it includes 2015 YY18 in Uranus' co-orbital region, and the two objects 2005 NP82 and 2016 YB13 near or in Jupiter's coorbital region to which  Ka`epaoka`awela, Jupiter's coorbital asteroid, is added for comparison. The polar Centaur 2010 CR140 that is also located near Jupiter's coorbital region was not included in the co-orbital group because the capture probability in Jupiter's coorbital region for nearly polar orbits is small \citep{NamouniMorais18c}. 

To the  high-inclination Centaurs, we add a fourth group with  the two known polar TNOs  2008 KV42 and (471325) 2011 KT19, and a fifth with the low-inclination Centaur (2060) Chiron. The distribution of the sample objects in the $xz$-plane ($y=0$) in terms of their semi-major axes, inclinations, perihelion and aphelion distances is shown in Figure 1. The first three columns of Table I list respectively the inclination $I$,  eccentricity $e$, semimajor axis $a$, and perihelion distance $q$. 

The Centaur and TNO nominal orbits and their equinoctial covariance matrices were obtained from the AstDys database\footnote{http://hamilton.dm.unipi.it/astdys \citep{Knezevic12}} for the Julian date 2458200.5 for all objects except  2005 NP82, 2010 CR140 and Chiron for which 2458400.5 was used (the Julian date 2457800.5 was used for Ka`epaoka`awela in Paper I.)  The planets' orbital elements  were obtained from NASA JPL's Horizons ephemeris system\footnote{http://ssd.jpl.nasa.gov \citep{Giorgini96}} for the same epochs. To apply our high-resolution statistical search for stable orbits, clones were generated for each object as in Paper I using the Cholesky method for multivariate normal distributions  in order to best reproduce the uncertainty amplitudes of their orbits \citep{cholesky}. 

In order to characterize the spread of a clone swarm in parameter space (or equivalently an orbit's uncertainty), we  make use of the generalized variance of the equinoctial covariance matrix $C$ \citep{Wilks32}. The latter is defined as the determinant of the covariance matrix ${\rm det}\,C$ and is used to estimate with a single number the spread of  multidimensional samples. As we seek to compare the swarm spreads of different objects to ascertain how the stable clone rate depends on the orbit's uncertainty, we define the relative generalized deviation of the orbit  RGD=$[{\rm det}\,C/(\Pi_{1\leq i \leq 6}e_i^2)]^{1/12}$ where $e_i$ are the six nominal equinoctial orbital elements. The generalized variance is divided by the product of all the orbital elements squared in order to obtain relative and not absolute spread estimates. The twelfth root is used to make the generalized deviation of the same order as a standard deviation since the covariance matrix elements are proportional to the square of standard deviation and $C$ is six-dimensional. The RGDs of the Centaur and TNO orbits  are listed in the fourth column of Table I.

The evolution of a Centaur clone back in time to $-4.5$\,Gyr when planet formation ended, was followed in the system composed of the four giant planets and the Sun whose mass was augmented by those of the inner Solar systemÕs planets. The full three-dimensional Galactic tide \citep{Heisler86} and relative inclination of the ecliptic and Galactic planes were taken into account. The Oort constants ($A = 15.3$ km\, s$^{-1}$ kpc$^{-1}$, $B = 11.9$ km\, s$^{-1}$ kpc$^{-1}$ ) and star density in the Solar neighbourhood ($\rho_0 = 0.119 M_\odot$\, pc$^{-3}$) are those derived from Gaia's first data release \citep{Bovy17,Widmark19}. Numerical integration of the five-body problem was carried out using the Bulirsch and Stoer algorithm with an error tolerance of $10^{-11}$ as in Paper I. Orbital evolution was monitored for collisions with the Sun, collisions with the planets, ejection from the Solar system, and reaching the inner 1 au semimajor axis boundary. No event at the inner boundary was registered for any of the Centaurs or TNOs. 

The clone number for each object was chosen with respect to the clone median lifetime determined from the first $10^4$ clone simulation. For a lifetime in excess of $10^7$ years, the number of 4.5\, Gyr-stable clones is relatively large and the statistics is determined with smaller clone sets  in order to reduce computational volume. The clone number varies  from $1.2\times 10^5$ for 2007 BP102, 2008 KV42 and  2011 KT19 to $3\times 10^5$ for 2016 FH13 and 2011 MM4. For smaller lifetimes and most objects, $10^6$ clones are used as in Paper I unless the stable clone rate is  $<10^{-5}$, then the clone number is increased to $2\times10^6$ (2010 CR40 and 2004 YH32). 

In the selected object sample, only Centaur (2060) Chiron is known to have exhibited cometary activity. Whereas the effect of non-gravitational 
forces is important in a comet's evolution, we do not include such processes in this study. Cometary activity is a complex phenomenon whose simplified modelling in the context of comet dynamics may produce less agreement with observations than a conservative approach \citep{Wiegert99}. As gravitational systems tend to the stable states available to them, the conservative simulation of Chiron's motion offers information on the stability of its current orbit and helps illustrate how the 4.5\,Gyr-stable orbit rate depends on the orbit's uncertainty.

\begin{table*}
\centering
\caption{Clone statistics at $-4.5$\,Gyr. Orbital inclination is denoted by $I$, eccentricity by $e$, semi-major axis by $a$ and perihelion by $q$. RGD stands for the relative generalized deviation of the Centaur's orbit. $T_0$ and $T_m$ are the minimum and median lifetimes. Orbital elements are not given to nominal orbit precision to avoid an overcrowded table. }
\label{table:1}
\begin{tabular}{lcccccccccccc}
\hline \hline
Asteroid  &$I$         & $e$ & $a$& $q$&RGD& Sample & $T_0$       & $T_m$        & Stable & Sun & Ejections & Planet  \\
               &$(^\circ)$&       & (au)& (au)& $(10^{-6})$& (10$^6$)&($10^4$yr)&($10^6$yr)  &  orbits         &collisions&&collisions\\
\hline
Polar Centaurs    & &&&&&&&&&&&\\
2010 FH92 & 62&0.76 &24.41&5.78&  0.58&1&3.09&2.19&15&242076&757843&66\\
2007 BP102 & 65&0.26 &23.98&17.75&  2.30 &0.12  &2.73&29.84&219&24151&95612&18\\
2010 CR140 & 75 &0.40&5.62&3.33&  0.64 &2&0.24&0.46 &14&894157&1105395&434\\
(144908) 2004 YH32& 79&0.56 &8.16&3.55& 0.24&2  & 0.78&0.68&17&873118&1126566&299\\
(518151) 2016 FH13 & 93&0.61 &24.55&9.46& 0.29 &0.3 &5.09&14.66&361&70951&228650&38\\
2014 JJ57& 96&0.29 &6.99&4.94& 1.11 &1  &0.12&0.83&24&435377&564313&286\\
2011 MM4 & 100&0.47 &21.13&11.14&  3.22  & 0.3&1.47&31.39&1295&69860&228776&69\\
(342842) 2008 YB3 & 105&0.44&11.62&6.50&   0.25&1  &4.46&2.69&129&299924&699699&248\\
2016 LS & 114&0.61 &13.34&5.23&  1.42 &1  &0.94& 2.43&106&254791&744858&245\\
\hline
Retrograde Centaurs     &&&&&&&&&&&&\\
2009 QY6   &137&0.83&12.47& 2.06&   0.37&1&3.80&3.26&41&446095&553637&227\\
1999 LE31    &152&0.47&8.13&4.34&  1.32 &1&0.60&2.62&69&301341&698136&454\\
2006 BZ8    &165&0.80&9.60&1.89&  0.36  &1&0.15&2.11&37&389485&610127&351\\
(330759) 2008 SO218   &  170 &0.56 &8.11&3.53&  0.32   &1&0.41&    2.56  &95&277445&721968&492\\
(434620) 2005 VD        & 173    &0.25 &6.67&4.99&   1.65  &1 & 0.08  &  2.41&91&275162&724175&572\\
\hline
Uranus coorbital   & &       &       &       &             &     &                    &               &                       &                 &                            &\\
2015 YY18     & 118 &0.83 &19.50&3.29&   0.27& 1&1.58&2.17 &28&375417&624418&137\\
Jupiter coorbitals               & &  &&&&&&&&&&\\
2005 NP82                        & 130 & 0.48&5.87&3.06&   0.21&1&0.11  & 1.33 & 17&487264 & 512286&433\\
2016 YB13                         & 140 & 0.41&5.46&3.23&  0.45 &1 &  0.64 &1.64 &31&443919&555453&597\\
(514107) Ka`epaoka`awela      &  163   &0.38&5.14&3.18&  0.45&  1&29.11&6.48   &  46            &    553811             &      445678                     & 465 \\
\hline
Polar TNOs   &&&&&&&&&&&&\\
2008 KV42                                        & 103&0.49 & 41.50&21.10& 6.91 & 0.12& 211.29& 811    &   9504       &18585&91885&26\\
(471325) 2011 KT19                         & 110 & 0.33&35.58&23.81& 0.64  &0.12&139.35& 511 &9359&20567&90041& 33 \\
\hline
Prograde Centaur   &&&&&&&&&&&&\\
(2060) Chiron                         & 7 & 0.38&13.66&8.45& 0.05  &1&0.26& 1.09&303&106731&891886&  1080\\
\hline \hline
\end{tabular}
\end{table*}

\section{Statistics}

The dynamical lifetimes and the clone statistics at $-4.5$\,Gyr are given in Table I. Regardless of orbit uncertainty (RGD),  the Centaur minimum lifetime scales as $10^3$ to $10^4$ years except those of Ka`epaoka`awela and the two TNOs which are $\sim$ 1\,Myr because of their privileged locations  in the Solar system's strongest mean motion resonance for the first one and beyond Neptune's orbit for the latter two. The high-inclination Centaur median lifetime $T_m$ ranges from 1 to 30\,Myr   years, and  is consistent with that of Centaurs generally \citep{TiscarenoMalhotra03,Horner04}. TNOs have  longer median lifetimes of 0.5\,Gyr  for 2001 KT19 \citep{Chen16} and 0.8\,Gyr for  2008 KV42. Surprisingly, 8 out of the 17 high-inclination Centaurs have median lifetimes that cluster around  2.4\, Myr with a standard deviation of 0.2\,Myr. They are 2015 YY18, 1999 LE31, 2006 BZ8, 2008 SO218, 2005 VD, 2010 FH92, 2008 YB3 and 2016 LS.  We discuss in Section 4, the possibility of mapping phase space using the median lifetime  to ascertain whether the 2.4\,Myr-lifetime Centaurs followed the same path in their evolution from the outer Solar system and were initially captured in similar conditions. The longest Centaur median lifetimes are found in the polar group with 15\, Myr for 2016 FH13 and 30\, Myr for both 2007 BP102 and 2011 MM4. Long median lifetimes seem to be correlated with large perihelia.  We searched for correlations between the ratio of the minimum and median lifetimes and the RGD without success  as we sought to ascertain whether the minimum lifetime scaled to the  median lifetime is an indicator of the level of orbit uncertainty.

Clone instability is mainly achieved by ejection and Sun collisions. Planet collisions are rare. Like the median lifetime, the relative statistics of unstable orbits show definite trends regardless of orbit uncertainty. Except for Ka`epaoka`awela, the fraction of ejected clones is always larger than that of Sun-colliding clones. Jupiter acts as a strong protector for Ka`epaoka`awela through its co-orbital resonance and as a strong perturber for all other Centaurs. In the polar group, the ejection rate increases  with semi-major axis  from $\sim$ 55 per cent to  80 per cent and the Sun-collision rate decreases accordingly with the sum of the two rates $\lesssim100$ percent. The opposite trend occurs in the retrograde group. The 2.4\,Myr-median lifetime group have ejection and Solar collision rates clustered around $70$ per cent and $30$ per cent respectively with a standard deviation of 5 per cent for both rates.

TNO clones are unstable largely through ejection at 75 per cent and through Solar collisions at 17 per cent whereas the largest rate of orbit ejection is by far Chiron's at 90 per cent as the orbit's low inclination maximizes the likelihood of disruption by close encounters with the planets.

\begin{figure*}
\begin{center}
{ 
\hspace*{-50mm}\includegraphics[width=280mm]{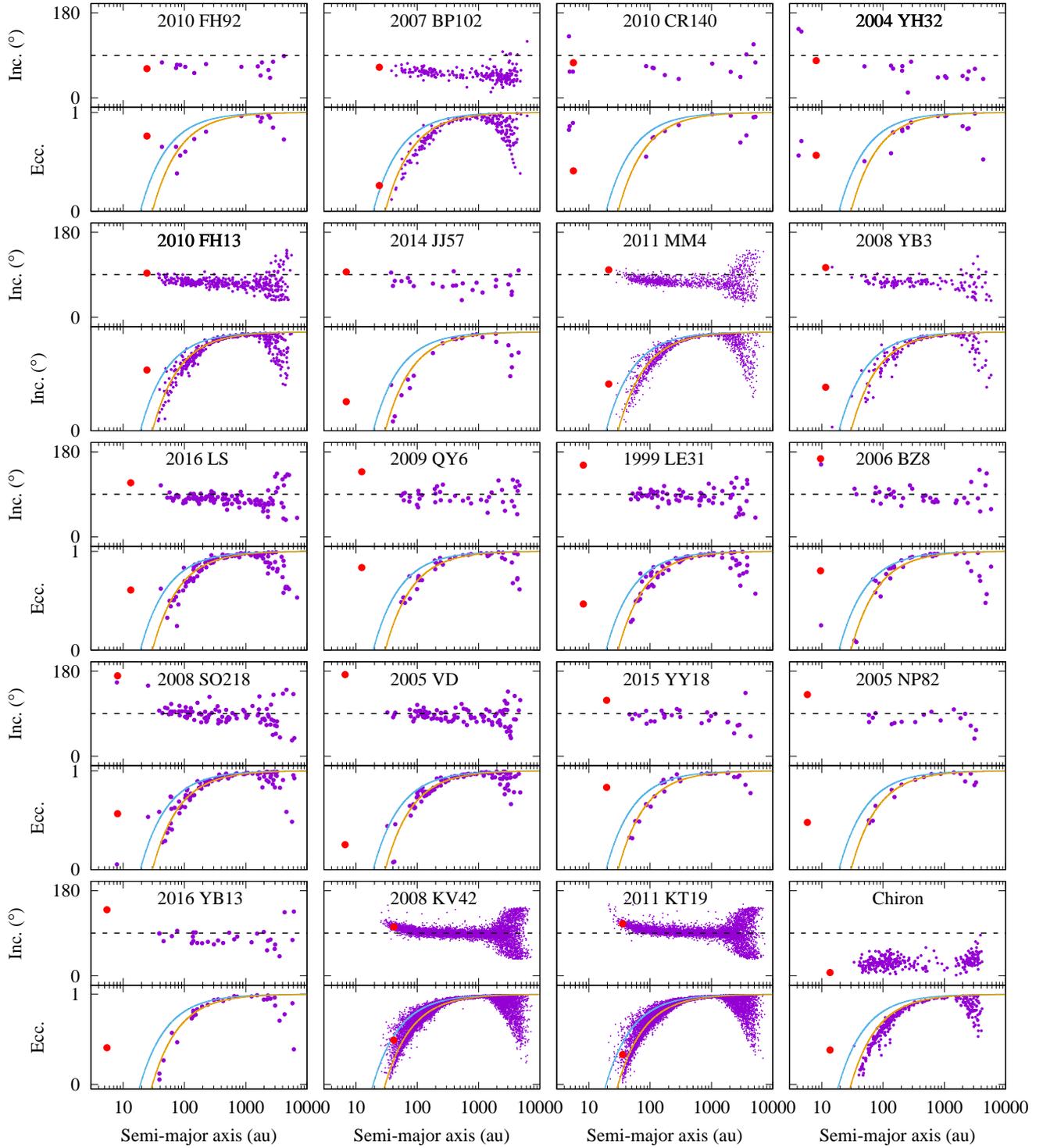}
}
\caption{Distribution of stable orbits in the eccentricity-semimajor axis and inclination-semimajor axis planes at $-4.5$\, Gyr. The  location of the object at the current epoch is indicated with a large red filled circle. The locus of intersection of the objects' perihelion with Uranus's and Neptune's orbits are shown with the top blue curve and the bottom orange curve respectively.}\label{f2}
\end{center}
\end{figure*}

Orbits that survive the age of the Solar system exist for all objects in the study: high-inclination Centaurs, the two TNOs as well as Chiron. The fraction of stable clones of high-inclination Centaurs ranges from 0.0007 per cent (2010 CR140) to 0.43 per cent (2011 MM4). In the 2.4\,Myr lifetime-group, the stable orbit rate is on average 0.007  per cent with a standard deviation of 0.004 per cent. The two TNOs have by far the largest stable orbit rate  of 8 per cent whereas Chiron stands at 0.03 per cent.

Large fractions of stable clones are naturally correlated with long median lifetimes.  They are  also related to the orbit's uncertainty. The most salient example is Chiron whose orbit is inclined by $7^\circ$ with respect to the planets and is located in a largely chaotic region as indicated by its 1\,Myr-median lifetime and its high planet collision rate. However, with 49 oppositions, it is by far the most observed object in the study.  Consequently, its orbit has the smallest RGD$=5\times 10^{-8}$ one to two order of magnitude smaller than the other objects in the study. 
This reflects on its stable orbit rate that is by far the highest among Centaurs with similar median lifetimes. 
{The two TNOs have RGDs that differ by an order of magnitude but the  median lifetime of the more uncertain orbit (2008 KV42) is about twice that of the less uncertain one (2011 KT19). The similar stable orbit rates of the two TNOs suggest that the long median lifetime of 2008 KV42, that signifies a less chaotic neighborhood than 2011 KT19's,  compensates its larger observational uncertainty than that of (2011) KT19.} In the 2.4\,Myr group, there is no clear relationship between RGD and stable orbit rate.

\begin{table}
\centering
\caption{Clone location breakdown at $-4.5$\,Gyr. ISS indicates the inner Solar system ($4\leq a ({\rm au})<5.2$),  1:1 the coorbital resonance, CEN centaurs ($5.2<a ({\rm au}) < 30$), KB the Kuiper belt ($30\leq a ({\rm au}) \leq 50$), SD the scattered disk ( $50< a ({\rm au}) \leq 10^3$,) and OC the Oort cloud ($a ({\rm au}) >10^3$). For the 1:1 resonance, the planet's initial identifies the asteroid's location.  }
\label{table:2}
\begin{tabular}{lcccccccc}
\hline \hline
Asteroid    & ISS & 1:1 & CEN  &KB&SD &OC\\
\hline
Polar Centaurs &&&&&&\\
2010 FH92 & 0 &  0&  0&1&7&7 \\
2007 BP102 & 0 &0&0&3&92&124\\
2010 CR140 & 1 & 2\,J & 0 &0 & 5&6 \\
(144908) 2004 YH32&2  &0  & 0 &1&7&7\\
(518151) 2016 FH13 & 0 &  0& 0 &12&203&146\\
2014 JJ57& 0 &  0&  0&3&13&8 \\
2011 MM4 & 0 & 1\,N& 2 &32&750&510\\
(342842) 2008 YB3 & 0 &  1\,N& 1 &1&80&46 \\
2016 LS &0  &  0&  0&1&66&39\\
\hline
Retrograde Centaurs     &&&&&&\\
2009 QY6   &0&0&0&0&24&17\\
1999 LE31    &0&0&0&1&46&22\\
2006 BZ8    &0&1\,S&0&2&23&11\\
(330759) 2008 SO218    &0& 0   & 2  & 3       & 58     &32\\
(434620) 2005 VD         & 0&   0    &  0  &   4   &     50       &37\\
\hline
Uranus coorbital               &&&&&&\\
2015 YY18 & 0 & 0&0 &1&19&8\\
Jupiter coorbitals               &&&&&&\\
2005 NP82                       &        0                   &     0     &           0    &  0             &11&6\\
2016 YB13                            &0    & 0    & 0   & 3    & 16        & 12 \\
(514107) Ka`epaoka`awela      &0                              & 27\,J          & 0               & 0      & 10    & 9\\
\hline
Polar TNOs   &&&&&&\\
2008 KV42              &0& 3\,N&6 &287& 6255&  2953\\
(471325) 2011 KT19&0& 8\,N    & 16     & 314   & 6256     & 2765 \\
\hline
Prograde Centaur  &&&&&&\\
(2060) Chiron        &0& 0&0&15 &193& 95\\

\hline \hline
\end{tabular}
\end{table}

\begin{figure*}
\begin{center}
{ 
\hspace*{-30mm}\includegraphics[width=230mm]{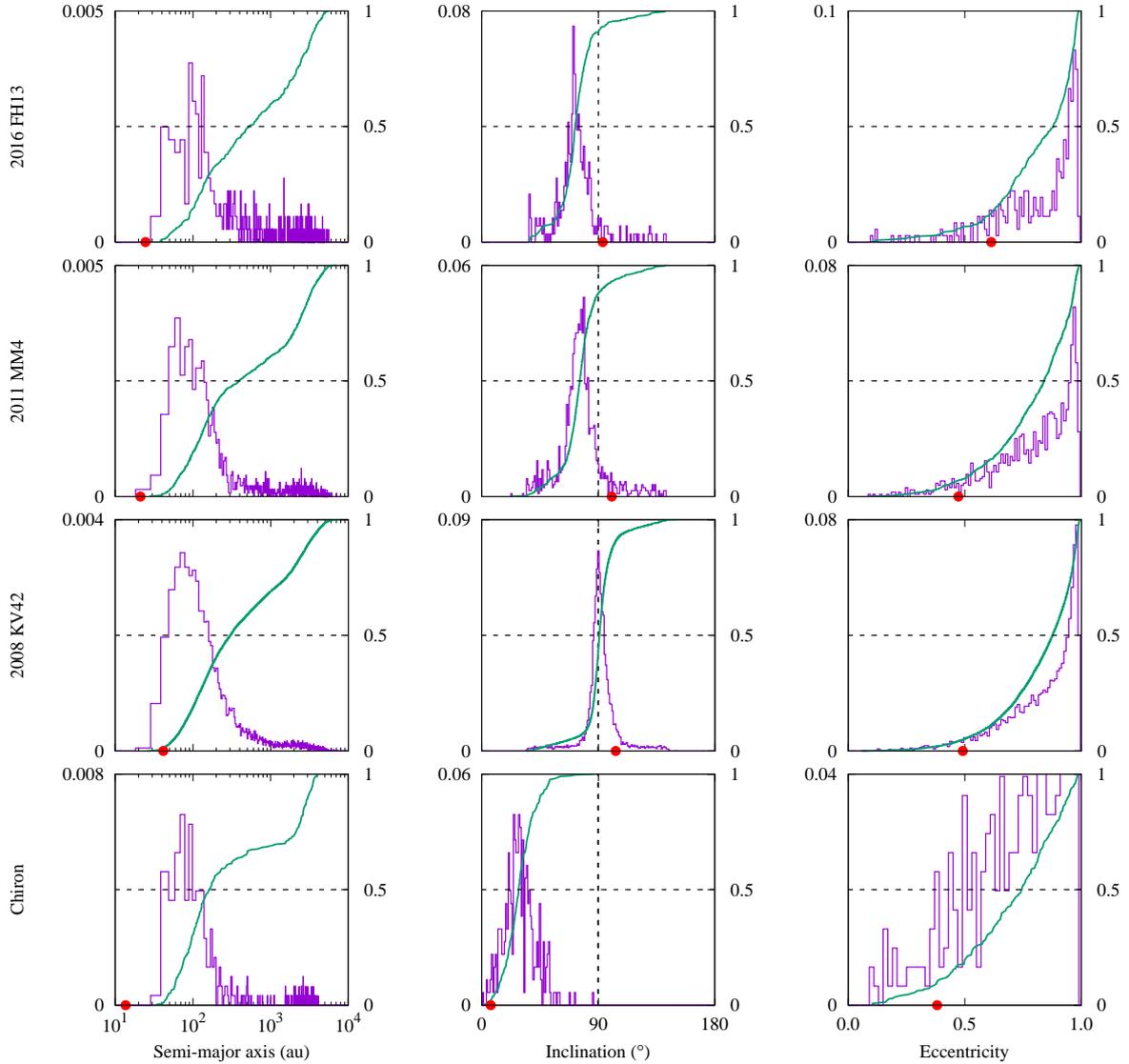}
}
\caption{Probability densities (purple curves, left scales) and empirical cumulative distribution functions (green curves, right scales) of some high stable orbit rate objects at $-4.5$ Gyr. The red filled circle on the horizontal axis of  each panel indicates nominal value of the corresponding orbital element at the current epoch.}\label{f3}
\end{center}
\end{figure*}

The distribution of stable clones {at -4.5\, Gyr} is shown in Figure 2 in terms of their  eccentricity $e$ , inclination $I$ and semi-major axis $a$  along with the current position of the object's orbit. For convenience, we divide the stable clones locations into six groups: the inner Solar system with $4\leq a ({\rm au})<5.2$, planet coorbitals, the Centaur region $5.2<a ({\rm au}) < 30$, the Kuiper belt region $30\leq a ({\rm au}) \leq 50$, the scattered disk region  $50< a ({\rm au}) \leq 10^3$, and the Oort cloud region  $a ({\rm au}) >10^3$. The $10^3$ au frontier delineates the region beyond which the Galactic tide's effects are important (Figure 2). The breakdown of the stable clones' location is given for all objects in Table 2. Except for Ka`epaoka`awela, the vast majority of stable clones are found in the scattered disk and Oort cloud regions. Clones in the co-orbital state, the inner Solar system, the Centaur region and the Kuiper belt region are rare.  The latter are present mainly for objects with large stable orbit rates such as those with large median lifetimes and Chiron whose orbit uncertainty level is low.  Below a current  inclination of $80^\circ$, the polar Centaurs' clones tend to be equally divided between the scattered disk and Oort cloud regions with a preference for the latter (2007 BP102). Above the threshold inclination, there is a clear preference of all objects for the scattered disk, including the two TNOs and Chiron whose clone ratio for the two regions  is 2. 

For objects with large stable orbit rates, the probability density and the empirical cumulative distribution function at $-4.5$  Gyr can show additional features regarding the distribution of the orbital elements. These functions are shown for  2016 FH13, 2011 MM4, 2008 KV 42 and Chiron in Figure 3. The semi-major axis, eccentricity and inclination  probability densities were binned in 10 au, 0.01 and $1^\circ$ intervals respectively. The semi-major axis distributions of the four objects peak between 70 au and 100 au with a smaller secondary peak in the inner Oort cloud. Probability density peaks show as inflection points on the (unbinned) cumulative distribution function. The inclination probability densities have a single peak between $70^\circ$ and  $90^\circ$ except for Chiron whose mean inclination is $29^\circ$.  The eccentricity probability densities peak between 0.8 and 0.9 and reflect for all objects the path followed by the clones. With perihelia near the orbits of the planets, clone eccentricity is invariably raised by planetary close encounters. If the clone reaches the Oort cloud region,  the Galactic tide lowers its eccentricity and increases its inclination in a secular periodic cycle \citep{Heisler86}.

\begin{figure}
\begin{center}
{ 
\hspace*{-15mm}\includegraphics[width=130mm]{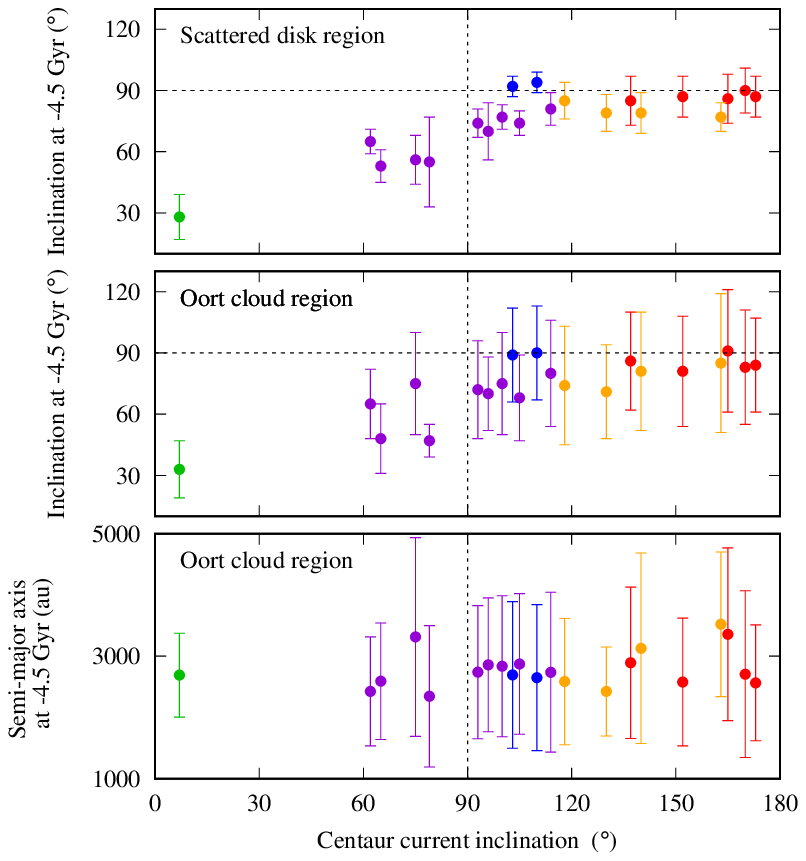}
}
\caption{Scattered disk clone inclinations and Oort cloud clone inclinations and semi-major axes at $-4.5$\,Gyr as functions of the Centaur's  inclination at the current epoch. The filled circles indicate the mean values, and the error bars the corresponding standard deviations. The color codes are those of Figure 1.}\label{f4}
\end{center}
\end{figure}

The stable orbits found in the scattered disk and Oort cloud regions all have high inclinations and are shown in Figure 4 in terms of their mean values and standard deviations for all objects as functions of the object's current inclination. Excluding Chiron, the only low-inclination Centaur, the mean inclination at $-4.5$\,Gyr in the scattered disk region is an increasing function of the inclination at the current epoch. It starts at $\sim 55^\circ$ for a current inclination of $\sim 60^\circ$ and reaches asymptotically to $ 90^\circ$ as current inclination increases to $180^\circ$. Standard deviation is approximately $10^\circ$. In the Oort cloud region a similar trend is found albeit with larger  mean inclination scatter and larger standard deviations explained naturally by the effect of the Galactic tide as can be seen in Figure 2. The parameter space structure containing the nearly-polar orbits was termed as the polar corridor in Paper I and  identified as the most favorable location for Centaurs that were captured by Jupiter and moved away from its orbit. 

Chiron's mean inclinations and standard deviations in the scattered disk and Oort cloud regions at $-4.5$\,Gyr respectively are $28^\circ\pm11^\circ$ and $33^\circ\pm 14^\circ$.   For all objects, the stable clones found in the scattered disk region have perihelia near Uranus's and Neptune's orbit (Figure 2). {In effect, as expected for Centaur-type motion, clones are scattered by the four giant planets toward the inner and outer parts of  the Solar system. Those sent toward the inner Solar Solar system are largely unstable as they are subjected to the perturbing effect of Jupiter and Saturn. Stable clones that are scattered to the outer Solar system are ultimately locked  to Uranus's or Neptune's perihelia unless they reach the Oort cloud region where the Galactic tide lowers their eccentricity and inclinations periodically.}   The stable clones found in the Oort cloud region share the same domain   with a semi-major axis centered around 2800\,au  of width 1200\,au (standard deviation)  (Figure 4).  This domain corresponds to the secondary peaks of the semi-major axis distributions shown in Figure 3. 

\section{Discussion}
In this work, we have applied a precise orbit determination method to determine the {possible} orbits of real high-inclination  Centaurs and TNOs at the end of planet formation. The statistical nature of this method implies that only probability distributions can be derived regarding the early orbits of the studied asteroids. This characterization is sufficient to point out the origin of the studied objects.  It enabled us for instance to show that all high-inclination Centaurs in the study had {probably} nearly-polar inclinations {4.5\,Gyr in the past. If high-inclination Centaurs had originated from the flat planetesimal disk of standard solar system formation theory, our statistical method would have found a majority of clones with relatively small inclinations near the common orbital plane of the outer planets for each of the objects in the study.}

{Provided that the clone number is sufficiently large, our orbit determination method does not depend on any ad hoc initial conditions such as those inherent to planetesimal disk relaxation models that research the origin of outer Solar system objects. In this respect, the method is not a model and its derived statistical orbit distributions for high-inclination Centaurs are robust and may not be fine-tuned  to change the outcome of the simulation.} The probability distributions can only be improved by further observations that reduce the uncertainty level of the studied objects' orbits.  That is why this method  can independently test origin theories of outer Solar system's objects by following back in time the motion of real Solar system objects. 

{To interpret precisely the finding that high-inclination Centaurs had nearly-polar orbits at the end of planet formation, it is useful to recall the current understanding of early Solar system structure \citep{Pfalzner15}. The Solar system started forming 4.6\,Gyr in the past. Sometime after a few Myr, giant planets formed and radially migrated while interacting with the planetesimal disk. The migration relaxation time is thought to have been about 1 to a few  10\,Myr whereas the full migration stage is thought to have lasted about a 100\,Myr to  500\,Myr at the end of which the planets reached their final orbits. The planets' orbital evolution in this stage is largely uncertain owing to the multitude of complex processes that operate in the first 100\,Myr of planet formation \citep{ribeiro20}. What is certain, however, is first, that the planetesimal disk before the migration stage as well as at the end of this stage had a small inclination dispersion in order to explain the very low inclination TNOs currently observed in the Kuiper belt (known as the cold population) \citep{Pfalzner15}. Secondly, the early planetesimal disk could not have extended beyond 30 au to 40 au \citep{Gomes04} to ensure that Neptune does not end up with a much larger final semi-major axis than the current one. Thus the early scattered disk  and Oort cloud were devoid of Solar system material. Our orbit determination method was run for 4.5\,Gyr in the past. The fact that the planets migrated towards the end of that period is not included in our calculation because the planets' evolution in the migration stage is largely uncertain but more importantly,  because it does not affect the Centaur orbits end states in any significant way. The reason lies in the fact that the bulk of stable Centaur clones that end up on nearly polar orbits do so in the first 1\,Gyr of the simulation. Therefore near 4\,Gyr in the past, the polar stable clones are already located far away in the scattered disk and the inner Oort cloud. Changes in the planets' semimajor axes over a few 10\, Myr do not affect such distant polar orbits significantly, as in reality, the planets' effect had long started to decay in the last billion years before migration even occurred. 

The high likelihood of nearly-polar 4.5\,Gyr-stable orbits for high inclinations Centaurs in regions devoid of Solar system material indicates that they did not belong to the nearly flat early planetesimal disk and have probably been captured from the interstellar medium. } The Sun's birth cluster of stars naturally provided a significant source of asteroids and comets for the Sun and the planets to capture during an epoch where gas in the Solar system and the interstellar medium was present to help seal temporarily trapped objects into permanent orbits around the Sun \citep{Fernandez00,Levison10,Brasser12b,Jilkova16,Hands19}. 

{Centaur capture from the interstellar medium has been invoked in a  recent work \citep{Kaib19} that examined disk relaxation models against the observed population of  Centaurs and TNOs in the outer solar system survey (OSSOS). Whereas that sample included only one high-inclination TNO (2008) KV42, present in our study,  and mainly low to moderate inclination orbits ($\leq 53^\circ$ with a median of $14^\circ$), the authors concluded that reproducing the high inclinations observed in their sample is not explained by the relaxation of the planetesimal disk with or without an additional hypothetical outer planet and that enrichment from the interstellar medium could solve this problem.  Our findings confirm that conclusion. Interestingly, another recent work on calibrating the OSSOS Centaur detections concluded that classical planetesimal disk models do not explain the abundance of high-inclination Centaurs and that `other sources may be needed' \citep{Nesvorny19}. }

Our finding that  the probable origin of high-inclination Centaurs is the  interstellar medium disagrees with earlier theories about Centaur origin being the relaxed primordial planetesimal disk \citep{Levison97,TiscarenoMalhotra03,Emelyanenko05,Disisto07,Brasser12}. Since our approach relies on a precise orbit determination method for real Centaurs that does not invoke any ad hoc initial conditions, our obtained orbit probability distributions are robust. The question is then what is the difference between the objects seen in relaxation models (with ad hoc initial conditions) and identified as possible Centaurs, and the Centaur clones in this study?  The main difference is related to the dynamical pathways followed by the two types of objects: the theoretical Centaurs  and the real Centaurs in our work. Disk relaxation models do not have sufficient dynamical resolution to reproduce the dynamics of real objects. Instead the primordial planetesimal disk is spread over a large spatial extent and integrated forward in time from the end of planet formation to the current epoch. There is no certainty that the identified dynamical pathways in such simulations are those of real Solar system objects beyond the  global  comparisons in parameter space (orbital elements, perihelia, Tisserand parameter) with entire populations of small bodies. 

When 4.5\,Gyr-stable orbits were found in Paper I for Ka`epaoka`awela with a majority of retrograde orbits in Jupiter's co-orbital region, it was concluded that Ka`epaoka`awela is likely of interstellar origin as no internal Solar system dynamical process could produce asteroid orbits with an inclination of $162^\circ$ at Jupiter's location at that early epoch. Ka`epaoka`awela could be a representative of a class of asteroids captured from the interstellar medium by the Sun and Jupiter owing to the strength of Jupiter's co-orbital resonance at large inclination that is responsible for shielding the asteroid from disruptive perturbations from the other planets  \citep{MoraisNamouni13a,MoraisNamouni16,NamouniMorais18b}. 

The existence of 4.5\,Gyr-stable orbits for high-inclination Centaurs, the two polar TNOs and Chiron widens the significance of that earlier finding in that stability over the age of the Solar system is possible even for Centaurs that are not protected by strong resonances. Furthermore, high inclination is not a prerequisite for Centaur stability over the age of the Solar system as Chiron's example demonstrates.

The definite trends that were identified from the statistics of unstable and stable orbits and the various relationships including the minimum and median lifetimes have the potential of shedding new light on the dynamical structure of phase space that classical local chaos indicators cannot. For instance the determination of the typical timescale for diverging orbits, classically used to identify dynamical chaos, would have concluded accurately only that all objects in the study are located in strongly chaotic regions of phase space. To gain more information regarding the existence of stable orbits, statistical methods are necessary to ascertain precisely  the possible pathways that Centaurs navigate on stable orbits in order to transfer from as far out as the inner Oort cloud to the giant planets' domain. In this regard, the surprising finding that 8 out of 17 high-inclination Centaurs have the same median lifetime of 2.4$\pm0.2$\,Myr could be an indication of a common pathway in phase space that was followed by objects captured under similar conditions or even in a single interstellar capture event. 

Directly identifying the locations of capture events depends on the stable orbit rate.  As explained in the Introduction, the million clone sample when projected on a single orbital space dimension is equivalent to sampling over 10 points within the error bar of the corresponding orbital element. With such  limited sampling, looking for common origins among different objects requires that either median lifetime is large or that the orbit is known with great certainty.  Both conditions guarantee a large stable orbit rate in order to favor clone clustering. An exception is given by Ka`eapoka`awela's location in Jupiter's strongest resonance where a majority of clones cluster on a single orbit instead of being scattered across a wide expanse from 50\,au to 4000\,au like all objects in this study.   {Identifying} common origin is likely to succeed at low inclination as such Centaurs have the lowest uncertainty levels owing to their extended observation history. In this respect, one of the surprising findings of this study is the stability of Chiron.  Although the simulation does not include non-gravitational forces, it shows that the current orbit is 4.5\,Gyr-stable and had an initial mean inclination  $\sim 30^\circ$.  {As the inclination dispersion of the planetesimal disk is believed to have been small before and after migration, there are  two possibilities for its origin. Either Chiron is an outlier that belonged to the planetesimal disk and whose cometary activity by some unknown mechanism increased its inclination far above the planetesimal disk's midplane, or it could be itself of interstellar origin.} Asteroid capture in the Sun's birth cluster does not necessarily favor objects whose orbits have or evolve to polar or high-inclination retrograde motion \citep{Hands19}. An astronomical illustration of the principle may be found in the distribution of the irregular satellites of the giant planets. Applying the high-resolution statistical stable orbit search to low inclination Centaurs is likely to shed light on the possible common capture events that occurred in the early Solar system.

\section*{Acknowledgments}
We are grateful to the reviewers for their comments that helped improve the clarity of the paper. The orbital search simulations were done at the M\'esocentre SIGAMM hosted at the Observatoire de la C\^ote d'Azur.  M.H.M. Morais research had financial support from S\~ao Paulo Research Foundation (FAPESP/2018/08620-1) and CNPq-Brazil (Pq2/304037/2018-4). This research was supported in part by FINEP and FAPESP through the computational resources provided by the Center for Scientific Computing (NCC/GridUNESP) of the S\~ao Paulo State University (UNESP).
 \bibliographystyle{mnras}

\bibliography{ms}

\end{document}